\documentclass[prb,twocolumn,floatfix,showpacs,superscriptaddress]{revtex4}
\usepackage{graphics}
\usepackage{dcolumn}
\usepackage{bm}
\usepackage{epsfig}

\begin{document}
\draft
\preprint{HEP/123-qed}
\title{Interplay of strongly correlated electrons and localized Ising moments in one-dimension}
\author{Chisa Hotta}
\affiliation{Kyoto Sangyo University, Department of Physics, Faculty of Science, Kyoto 603-8555, Japan}
\date{December 3, 2009}
\begin{abstract}
We study the ground state properties of the one-dimensional quarter-filled strongly correlated electronic chain 
coupled by $J$ to another chain of antiferromagnetic Ising moments. 
We focus on the case where the large Coulomb interactions localize the charges on every other site. 
Both the electronic spins and the Ising moments interact 
antiferromagnetically within each chain by $J_{\rm eff}$ and $J'$, respectively. 
Since the number of electrons is half as that of the Ising moments 
the period of magnetic correlation of these two chains are incommensurate. 
In the presence of $J$, the frustration of $J_{\rm eff}$ and $J'$ arises,  
which may lead the system to the intriguing magneto-electric effect. 
\end{abstract}
\maketitle
\narrowtext
\newcommand{\Beqn}{\begin{equation}}
\newcommand{\Eeqn}{\end{equation}}
\newcommand{\Beqna}{\begin{eqnarray}}
\newcommand{\Eeqna}{\end{eqnarray}}
\newcommand{\nonn}{\nonumber \\}
\newcommand{\expo}{{\rm e}}
\newcommand{\imag}{{\rm i}}
\newcommand{\nsig}{\bar{\sigma}}
\newcommand{\expct}[1]{\langle #1 \rangle}
\newcommand{\cket}[1]{| #1 \rangle}
\newcommand{\bra}[1]{\langle #1 |}
\newcommand{\ag}[1]{$\langle {\rm #1} \rangle$}
\newcommand{\wtilde}[1]{\widetilde{#1}}
\newcommand{\dagg}{^{\:\dagger}}
\newcommand{\wdagg}{^{\;\dagger}}
\newcommand{\ndagg}{^{\dagger}}
\renewcommand{\Re}{{\rm Re}}
\newcommand{\dwn}{\downarrow}
\newcommand{\up}{\uparrow}
\section{Introduction}
The electrons coupled to localized spins have been a long studied issue 
from Kondo chains in heavy fermionic systems \cite{aeppli}, 
double exchange systems(DEX) in manganites\cite{dagotto99} to $\pi$-d systems of molecular solids
\cite{pd,kobayashi00,sugimoto04,uji01}. 
Particularly, the negative giant magnetoresistance effect in DEX system  
had been a highlight which provided a mechanism of tuning the electronic degrees of freedom 
by the magnetic field\cite{dagotto99,kubo72}. 
The author and the co-workers tried to figure out another possible mechanism of 
the negative magnetoresistance effect by considering the Kondo chains including the Coulomb interaction terms, 
which we call an extended Kondo lattice model(EKLM)\cite{chisa}. 
The study on EKLM was carried out with in mind the experimental findings of 
the giant negative magnetoresistance in the one-dimensional (1D) organic solid, 
TPP[Fe(Pc)(CN)$_2$]$_2$\cite{hanasaki00}. 
This so-called phthalocyanine compound is a quarter-filled $\pi$-electronic system 
which includes a localized Fe d-spin on the same TPP-molecule\cite{tajima_chem}. 
The $\pi$-electrons have a charge order, which is stabilized by the $\pi$-d interaction. 
However, the EKLM model turned out to be different from the experimental situation 
mainly in two points; 
experimentally, the localized moment has a large Ising anisotropy as an artifact of the 
spin-orbit coupling\cite{hanasaki03}, and between them, there possibly exists a direct interaction. 
These factors are not considered in the EKLM where the localized SU(2) spins do not have 
the direct interaction with each other. 
We revisit this problem by taking into account these points. 
\par
The present paper deals with two chains which have electrons and Ising moments, respectively. 
We focus on the strong coupling case where the electrons are localized (insulator). 
Then the system is essentially regarded as two spin chains which have antiferromagnetic intra-chain 
correlation with different periodicity. 
The introduction of the coupling between the two chains lead to the "magnetic frustration". 
The paper is organized as follows; Sec.II gives the model and details of 
the numerical analyses, and Sec.III the ground state phase diagram. 
Sec.IV is devoted to the clarification of the magnetic properties of the electrons, 
and finally the external magnetic field is introduced in Sec.V. 
After all we reach the anticipation that the "magnetic frustration" 
provide a more realistic example of the magneto-electric effect. 
%
\section{Models and methods}
We consider 1D Ising spin chain and electronic chain which are connected via Hund coupling $-J<0$ 
at each lattice site. 
Within the chain, the Ising moments interact antiferromagnetically by $J'>0$, 
while the electrons have strong on-site and inter-site Coulomb interactions, 
$U$ and $V$, respectively. 
The Hamiltonian reads, 
\begin{eqnarray}
\hspace*{-15mm}
&& {\cal H}={\cal H}_{\rm hubb} + {\cal H}_{\rm ising} + {\cal H}_{J}
\nonumber \\
&& {\cal H}_{\rm hubb} = -\!\sum_{\expct{ij}\sigma}\!\!\big(t c_{i\sigma}^\dagger c_{j\sigma}\! + {\rm h.c.}\big)
          +\!\sum_{\expct{ij}}V n_i n_j 
          + \sum_j U n_{j\uparrow}n_{j\downarrow}
\nonumber \\
&& {\cal H}_{\rm ising} = \sum_{\expct{ij}} J' S_i^z S_j^z
\nonumber \\
&& {\cal H}_{J} = - \sum_j J S_j^z s_j^z.
\label{ham}
\end{eqnarray}
Here, the operators $c_{j\sigma}$, $n_j$, and $s_j^z$ denote 
the annihilation, number, and the $z$-component of the spin at $j$-th site for electrons, respectively. 
We define the magnetization of electrons as $M_{\rm el}=\sum_j s_j^z$, and the magnetic density, $m_{\rm el}=M_{\rm el}/N$, where $N$ is the system size. 
The z-component of the localized Ising moment is represented by $S_j^z=\pm 1/2$, 
and $\expct{i j}$ denotes the nearest-neighbor pair sites. 
We focus on the case of quarter-filling of electrons where the $4k_F$-instability 
of the charge degrees of freedom is significantly large. 
\par
At $J=0$, each of the decoupled chains has well defined ground state; 
the Ising chains is a gapped spin system with N\'eel order, which is described as an antiferromagnetism of 
$4k_F$-periodicity, ($\up\dwn\up\dwn\cdots$), where $k_F$ is defined by the filling factor of electrons. 
On the other hand, the weak-coupling region of the electronic chain is a Tomonaga-Luttinger liquid (TLL). 
At quarter-filling, it undergoes a transition into a ordered state with a finite charge gap \cite{mila93,sano94,tsuchi00} 
when the interactions become as large as $U\gtrsim 4t$ and $V\gtrsim 2t$. 
In this ordered state at large $U$ and $V$, the electrons localize on every other site. 
The effective antiferromagnetic interaction between the spins of these electrons is given as, $J^0_{\rm eff}=t^4/UV^2$, 
which leads to the correlation characterized as a $2k_F$-spin-density-wave (SDW). 
Once $J$ is switched on, there arises a magnetic frustration between these two orders. 
Namely, the antiferromagnetic configuration of Ising spins favors the electronic spins to align ferromagnetically 
in order to gain the energy by $J$, which is incompatible with the antiferromagnetic correlation 
within the electronic chain by $J_{\rm eff}$. 
Such situation is given schematically in Fig.~\ref{f1}(a). 
\par
Another reference system is the extended Kondo lattice model (EKLM) in Ref.[8]. 
This model is realized if we replace the interaction of Eq.(\ref{ham}) 
between electrons and localized spins by that of the SU(2) symmetry, and take $J'=0$. 
The role of the quantum fluctuation of the localized spins shall be discussed shortly 
by comparing these two models. 
\par
In order to clarify the ground state of this system, we perform the density matrix 
renormalization group analysis (DMRG) \cite{white}. 
Usually, the Ising anisotropy of spin degrees of freedom works as disadvantage 
in optimizing variationally the restricted basis of the finite system. 
This is because the quantum fluctuation which usually works to improve the selection of 
the basis in the local update processes is present only in the electron hopping term. 
Therefore, by the straightforward application of DMRG to Eq.(\ref{ham}), 
one finds difficulty in selecting the optimized configuration of Ising moments. 
Instead, we consider the extended Hubbard chain in the presence of periodic field, 
described by the Hamiltonian, 
\begin{equation}
{\cal H}_{\rm el}= {\cal H}_{\rm hubb} - \sum_j W_j s_j^z. 
\label{elham}
\end{equation}
Here, $W_j=J \langle S_j^z\rangle$, is the on-site "magnetic field" created by the localized Ising moments. 
The calculations are given as follows; 
(i) assume several different configurations of Ising moments, 
(ii) calculate ${\cal H}_{\rm el}$ in DMRG under the potentials from each of these configurations at several 
system size $N$, 
and obtain the energy per site in the bulk limit by the finite size scaling analysis, 
(iii) add to $\expct{{\cal H}_{\rm el}}$ the interaction energy of Ising moments, 
$E_{\rm ising}\equiv \langle {\cal H}_{\rm ising}\rangle$, 
and get the lowest energy state as a function of $J'$. 
As for the Ising spin configuration in (i), 
we consider up to periodicity over sixteen sites, and it turns out that 
the states which have two- or four-fold periodicity give the lowest energies. 
This holds even in the non-interacting case, except at $J/t<0.005$ 
where a small ferrimagnetic region is found. 
We thus mainly focus on ($\up\up\up\up$), ($\up\dwn\up\dwn$), ($\up\up\dwn\dwn$), 
and ($\up\up\dwn\up$)-configurations, 
where $\up$ and $\dwn$ correspond to up and down spin Ising spins, respectively. 
Figures~\ref{f1}(a)-\ref{f1}(d) show the representative configuration of 
Ising moments combined with 
the unpolarized AF ($m_{\rm el}=0$) and fully polarized F ($m_{\rm el}$=0.25) electronic states. 
In the ground state phase diagrams of Sec.III we consider only the AF- and F-electronic states, 
which allows for the systematic finite size scaling in (ii). 
We confirmed in advance that the partial ferromagnetic states with ferrimagnetic Ising moments, 
e.g. ($\up\up\up\dwn$), do not appear as a ground state in the focus parameter region. 
In Sec.IV, we also include the Ising configurations up to 32-fold periodicity and 
calculate the $m_{\rm el}$-dependence by fixing $N$, 
in order to examine the effect of magnetic field. 
\par
In the finite system with open boundary condition, the charges have the largest density 
at both edge sites. If the charge order is dominant the charges tend to align 
in every other sites from both ends, and a calculation on even-$N$ yields 
a kink structure at the system center, e.g. for $N=8$ we 
find ($\bullet\circ\bullet\circ\circ\bullet\circ \:\bullet$), where $\bullet$ and $\circ$ are the charge rich and poor 
sites, respectively. 
In such case, the amplitude of the charge density is gradually suppressed towards the center site. 
If we take the odd number of sites, this kink disappears, while the electon number deviates by one 
from that of the quarter-filling\cite{shibata}. 
The results of the finite size scaling of ${\cal H}_{\rm el}$ up to $N=97$ for 
($\up\dwn\dwn\up$)- and ($\up\up\dwn\dwn$)-type of Ising moments 
(which are the same configuration in the bulk limit) 
with both the odd- and even-$N$ are presented in Fig.~\ref{f1}(b). 
At $N\rightarrow \infty$, the energy of all cases coincides within $<10^{-5}t$. 
Thus the energy of the bulk limit is safely obtained and the treatment (iii) is performed. 
By introducing finite $J'$, the energy density of the $(\up\dwn\up\dwn)$ and $(\up\up\up\up)$-state shifts by 
$-J'/4$ and $J'/4$, respectively, whereas that of the $(\up\up\dwn\dwn)$-state does not change. 
The lowest energy state among the calculated candidates are thus obtained as 
functions of $U$, $V$, $J$ and $J'$ at fixed $t=1$. 
\section{Phase diagram}
\subsection{Comparison of the ground state of the Ising and SU(2) localized moments at $J'=0$}
We first present in Fig.~\ref{f3}(a) the phase diagram of the Ising moments 
at $J'=0$, i.e. when the direct interaction between Ising moments is absent. 
This diagram is to be compared with the case of the EKLM which has the SU(2)-symmetry of localized moments\cite{chisa}. 
At small $J$, the ($\up\up\dwn\dwn$)-Ising moments couple with the TLL with 2$k_F$-SDW correlation, 
which undergoes a phase transition into the ($\up\up\up\up$)-Ising moments with fully polarized electrons. 
This situation has good correspondence with the EKLM; 
the present ($\up\up\dwn\dwn$)- and ($\up\up\up\up$)-Ising spin configurations are 
interpreted as the paramagnetic and the ferromagnetic state of the SU(2) spins in EKLM, respectively. 
The phase boundary shifts in the Ising case to about four times larger value from the SU(2) one. 
To understand this, let us consider the non-interaction case, $U\!=\!V\!=\!0$; 
in the EKLM , the ferromagnetic state at large $J$ is described by the formation of singlet pairs of 
electrons and localized moments, 
and the rest of the moments (the electron number is half the number of localized moments) 
are fully polarized by the hopping of singlets\cite{tsunetsugu93}. 
One can roughly approximate the energy of the paramagnetic and the ferromagnetic states as, 
$e_{\rm para}\sim -4\sqrt{2}t+\frac{J}{2}$ and $e_{\rm ferro}\sim -4t-\frac{J}{4}$, respectively. 
At $J\sim 2t$, tne phase transition takes place. 
On the other hand, the energy of the corresponding Ising-case is estimated by the 
modification of band structure under the periodic potential of Ising spins, 
which yields, $e(\up\up\dwn\dwn)\sim -\sqrt{\frac{J^2}{16}+2t+\sqrt{\frac{J^2}{4}+2t}}$ and 
$e(\up\up\up\up)\sim -4t-\frac{J}{4}$, respectively. 
The level crossing occurs approximately at $J/t\sim 6$, which is consistent with the phase diagram. 
Here, we note that the ($\up\up\dwn\dwn$)-AF state at small $U$ and $V$ 
is a band-insulator, because the four-fold periodicity 
leads to the formation of four isolated bands, where the lowest band is completely filled. 
The above discussions indicates that the lack of quantum fluctuation of the Ising moments 
leads to $e(\up\up\dwn\dwn) < e_{\rm para}$, 
which originates mainly from the large energy gain of $J$-term in the $(\up\up\dwn\dwn)$-state. 
Thus the phase boundary is pushed towards larger $J/t$ in the present Ising system. 
\begin{figure}[tbp]
\includegraphics[width=8.5cm]{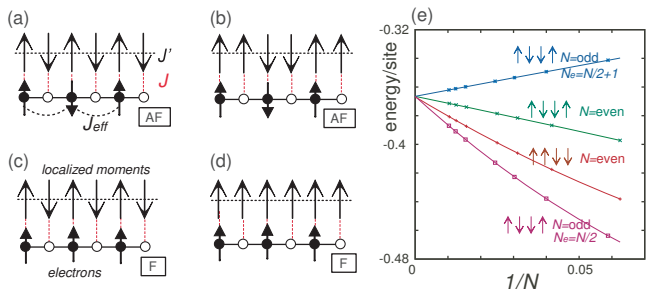}
\caption{(a)-(d) Representative configurations of localized moments and electrons in the 
strong coupling region. 
(b) Energy per site as a function of $1/N$ at $U/t=8$, $V/t=4$ and $J/t=1$ 
with ($\up\dwn\dwn\up$)- and ($\up\up\dwn\dwn$)- Ising spin configurations. 
System size scaling is given for four different series; odd $N$ with $N_e=N/2$, 
$N_e=N/2+1$, and even $N$. 
} 
\label{f1}
\end{figure} 
\begin{figure}[tbp]
\includegraphics[width=8cm]{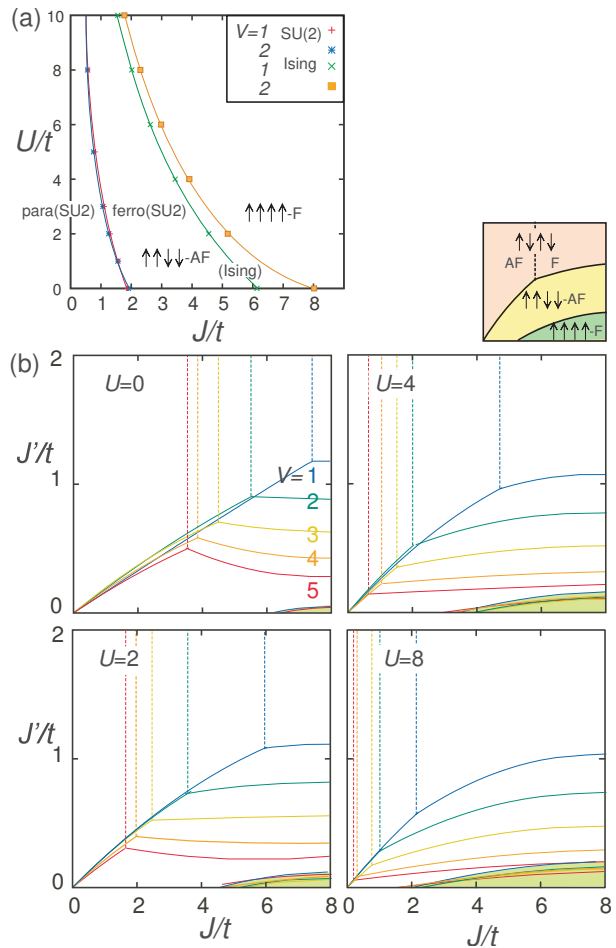}
\caption{
(a) Phase diagram at $J'=0$ on the plane of $U/t$ and $J/t$ for $V/t=1,2$. 
Phase boundary of the extended Kondo lattice model in Ref.[8] is plotted together 
for comparison. 
(b) Phase diagrams on the plane of $J/t$ and $J'/t$ for 
all sets of combinations among $U/t=0,2,4,8$ and $V/t=1,2,3,4,5$. 
As shown schematically in the inset, the diagrams are separated into four regions, 
($\up\up\dwn\dwn$)-AF, ($\up\dwn\up\dwn$)-AF, ($\up\dwn\up\dwn$)-F, and ($\up\up\up\up$)-F, 
which are shown schematically in Figs. 1(a)-(d). 
Here, AF and F denote the unpolarized and fully polarized electron spins, 
$m_{\rm el}$=0 and 0.25, respectively. 
} 
\label{f2}
\end{figure} 
\subsection{Phase diagram of localized Ising moments}
\begin{figure}[t]
\includegraphics[width=8.5cm]{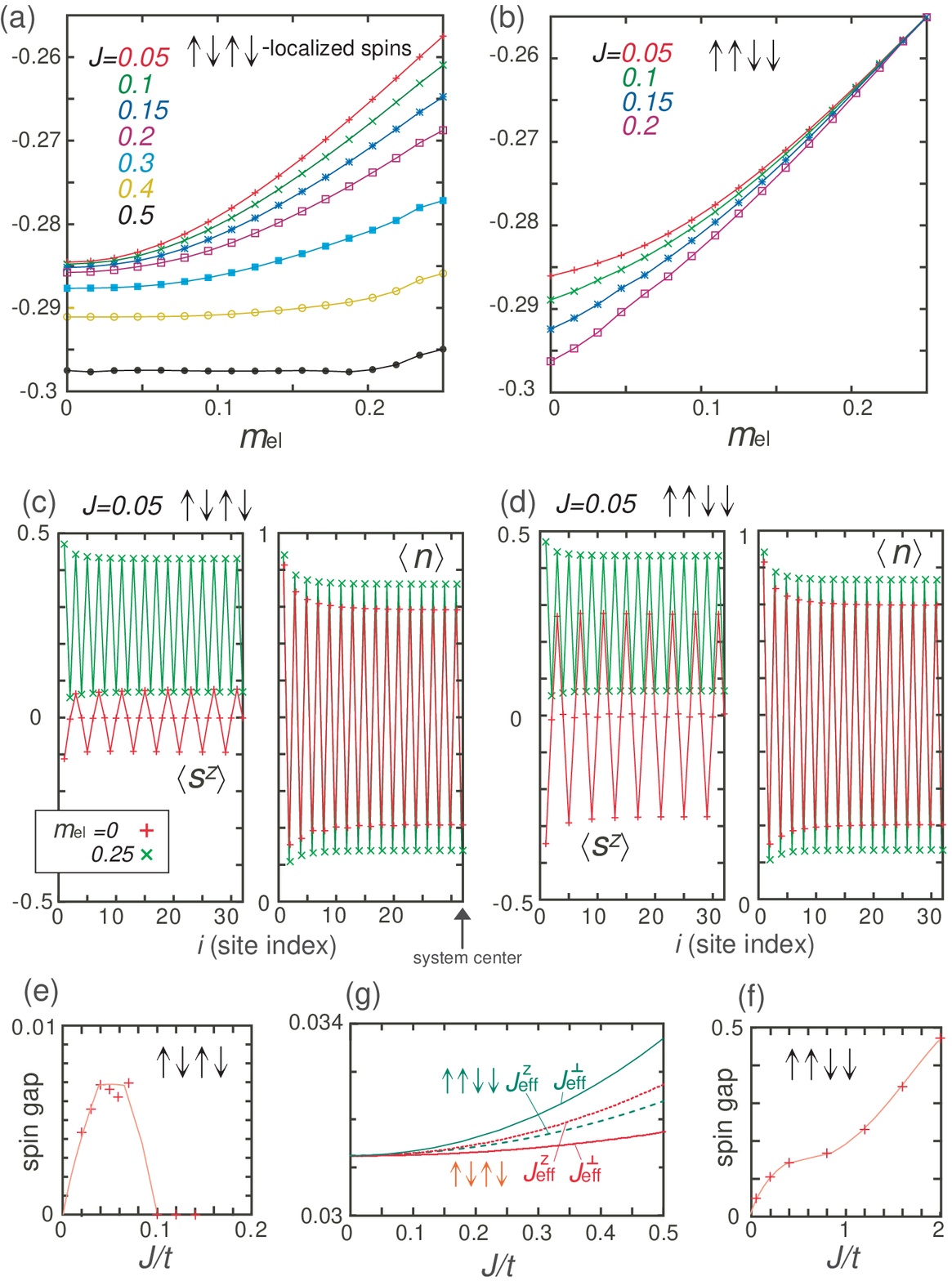}
\caption{Panels (a) and (b) are the $m_{\rm el}$-dependence of energy per site, $E(m_{\rm el})/N$, 
at $N=64$ with ($\up\dwn\up\dwn$)- ($\up\up\dwn\dwn$)-Ising spin configuration, respectively, 
at $U/t=8$ and $V/t=4$ for various choices of $J/t$.
Panels (c) and (d) are the local charge and spin densities, $\expct{n_i}$ and $\expct{s^z_i}$, 
which correspond to (a) and (b), respectively, at $J/t=0.05$. 
Panels (e) (f) are the spin gap in the bulk limit as a function of $J/t$ at $U/t=8$ and $V/t=4$ 
for ($\up\dwn\up\dwn$)- and ($\up\up\dwn\dwn$)-cases. 
Panel (g) give the effective spin-spin interaction $J_{\rm eff}$ given in Eqs.(\protect\ref{jeff1010}) and  (\protect\ref{jeff1100})
as a function of $J/t$ at $U/t=8$ and $V/t=4$. 
} 
\label{f3}
\end{figure}

In the next step, we include the $J'$-term and 
find that the ground state undergoes a phase transition into another magnetic state. 
Figure~\ref{f3}(b) shows the phase diagram classified by the configuration of localized moments 
on the plane of $J$ and $J'$ at several fixed values of $U$ and $V$ in unit of $t$. 
The interaction energies of the Ising moments for 
the configurations ($\up\dwn\up\dwn$), ($\up\up\dwn\dwn$), and ($\up\up\up\up$) are 
$E_{\rm ising}/N=-J'/4$, 0, and $J'/4$, respectively. 
Therefore, by the introduction of $J'$, the ($\up\dwn\up\dwn$)-state which gains the 
energy replaces the others. 
The boundary of the ($\up\dwn\up\dwn$)- and ($\up\up\dwn\dwn$)- phases are approximately 
given as $J \sim 4J'$, which is understood by the comparison of  
the magnetic energy of the ($\up\dwn\up\dwn$)-AF and ($\up\up\dwn\dwn$)-AF states in Fig.~\ref{f1}(a); 
the Ising moments have interaction energies, $E_{\rm ising}/N=J'/4$ and $0$, respectively. 
As for the $J$-term, the straightforward estimation gives, $E_{J}/N=0$($\up\dwn\up\dwn$-AF) and $J/8$($\up\up\dwn\dwn$-AF).  
However, as we see in Sec.III (Fig.~\ref{f3}(d) at $m_{\rm el}=0$), 
the amplitude of electronic spin moment in the ($\up\up\dwn\dwn$)-state is $\expct{s_z}\sim 0.25$, 
which is suppressed to about half the expected value. 
This is presumably because the electrons are relatively delocalized to the neighboring site 
(since the neighboring site has the same potential from the ($\up\up\dwn\dwn$)-potential). 
The resultant $E_J/N$ of ($\up\up\dwn\dwn$)-AF is $J/16$, and after the comparison of $E_{\rm ising}+E_J$ in both states, 
the phase boundary falls on $J\simeq 4J'$. 
\par
The phase boundaries are influenced by the electronic interactions as well. 
By comparing the phase diagrams we find that $U$ stabilizes the ($\up\up\up\up$)-F state. 
This is because the exclusion of double occupancy due to $U$ favors magnetism. 
On the other hand, the ($\up\dwn\up\dwn$)-F state ($m_{\rm el}=0.25$) is stabilized by both $U$ and $V$. 
When the electronic spins are the fully polarized, 
$J$ works as potentials to pin the electrons on every other sites, which favors charge order. 
The ($\up\dwn\up\dwn$)-F with large $U$ and $V$ is the typical example. 
Thus $U$, $V$, and $J$ cooperatively stabilizes the charge ordered state. 
Such effect of Coulomb interactions are consistent with what we find in the EKLM\cite{chisa}. 
\par
The ($\up\dwn\up\dwn$)- and ($\up\up\dwn\dwn$)-phases at $U=8$ given in Fig. \ref{f2}(b) are insulators. 
We confirmed this by the finite size scaling analysis on the charge gap for several choices of parameters. 
The region of $U/t \gtrsim 4$ and $V/t \gtrsim 2$ is an insulator at $J$=0\cite{mila93}. 
The introduction of $J\ne 0$ stabilizes the insulating phase for both cases, ($\up\dwn\up\dwn$) and ($\up\up\dwn\dwn$). 
Therefore, our discussion regarding the magnetic properties of the insulating state 
in the next section is safely carried out. 
%
%
\section{Magnetic properties}
\subsection{Competing magnetic orders}
The ground state phase diagrams are basically dominated by 
the ($\up\dwn\up\dwn$)- and ($\up\up\dwn\dwn$)-configurations of Ising moments. 
So far, for each of these cases, the spin-unpolarized $m_{\rm el}$=0 (AF) and 
fully spin-polarized $m_{\rm el}$=0.25 (F) electronic states are examined. 
In the next step, we calculate the $m_{\rm el}$-dependence of energy of the electronic state by DMRG. 
Here, we focus on the large $U$ and $V$-region where the system is a charge ordered insulator. 
The nontrivial competition among $J$, $J'$ and $J_{\rm eff}$ we discussed in Sec.II 
shall be examined explicitly, 
where $J_{\rm eff}$ is the effective interaction between electronic spins localized on every other site. 
Figures~\ref{f3}(a) and \ref{f3}(b) show the energy per site, 
$E(m_{\rm el})/N$ under the above mentioned two different configurations 
for several choices of $J$ at $N=64$, $U/t=8$ and $V/t=4$. 
In the ($\up\dwn\up\dwn$)-state, the functional form of $E(m_{\rm el})$ is flattened as $J$ increases. 
Contrastingly, $E(m_{\rm el})$ of the ($\up\up\dwn\dwn$)-state 
becomes a more rapid increasing function at larger $J$. 
\par
The charge and electronic spin density for $m_{\rm el}$=0 and 0.25 under 
($\up\dwn\up\dwn$)- and ($\up\up\dwn\dwn$)-configuration are given 
in Figs.~\ref{f3}(c) and \ref{f3}(d). 
The spin densities differ significantly between the two figures; 
at $m_{\rm el}=0$ the ($\up\dwn\up\dwn$)-state has small amplitude of $\langle s_z\rangle$ 
compared to that of the ($\up\up\dwn\dwn$)-state. 
The former has the $4k_F$-antiferromagnetic spin correlation on the Ising chain which does not fit to that of the $2k_F$-one along the electronic chain. 
Therefore, the spin density is rather suppressed to cope with $J$ which connects the frustrating 
two different spin correlations. 
In contrast, the latter ($\up\up\dwn\dwn$)-state  has commensurate 2$k_F$-antiferromagnetic correlation 
along both the electronic and Ising spin chains. The cooperation of the two correlations enhances the amplitude of the spin moment. 
When the electrons become fully polarized at $m_{\rm el}$=0.25, the magnetic frustration in the ($\up\dwn\up\dwn$)-state is resolved 
so that the difference between the two configurations becomes almost negligible.  
\par
The spin gap is shown in Figs.\ref{f3}(e) and \ref{f3}(f) as a function of $J/t$ at $U/t=8$ and $V/t=4$, 
which is obtained after the extrapolation to the bulk limit. 
Both cases starts from a gapless state at $J$=0. 
Under ($\up\dwn\up\dwn$)-configuration, a small gap opens first and then closes again already at extremely small $J$.  
As for the ($\up\up\dwn\dwn$)-case the spin gap continues to increase as a function of $J$. 
These results are consistent with the findings in Figs.~\ref{f3}(a) and \ref{f3}(b). 
\par
As we discussed in Sec.III, the charge order is stabilized by the polarization of the electronic spins 
in both ($\up\dwn\up\dwn$) and ($\up\up\dwn\dwn$)-cases. 
Actually, the amplitude of $\expct{n_i}$ increases with increasing $m_{\rm el}$. 
It is interesting to find that $\expct{n_i}$ does not seem to differ 
between the ($\up\dwn\up\dwn$)- and ($\up\up\dwn\dwn$)-cases even at small $m_{\rm el}$ 
where the magnetic properties of the two significantly differ. 
Namely, at quarter-filling the charge degrees of freedom is rather decoupled to the spin degrees of freedom. 
This does not hold off-quarter-filling which we discuss in Sec. V. 
%
\begin{figure}[t]
\includegraphics[width=7cm]{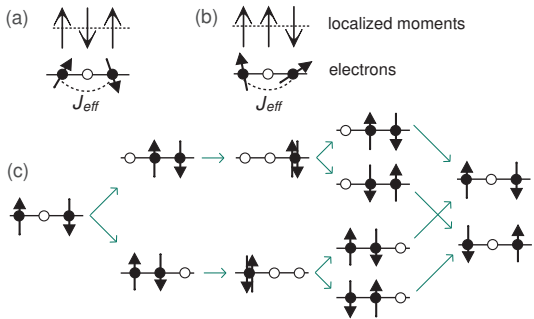}
\caption{The representative charge ordered state at large $U/t$ and $V/t$ for 
(a)($\up\dwn\up\dwn$)- and (b)($\up\up\dwn\dwn$)-configurations, 
where the Ising moments work as internal uniform and staggard field to the electronic spins, respectively. 
Panel (c) is the fourth order perturbative process by the hopping of electrons. 
The processes themselves are common between (a) and (b) 
while the energy of each state in the processes differ due to different configuration of Ising moments. 
} 
\label{f4}
\end{figure} 
\subsection{Strong coupling approach}
In order to understand the nature of such spin degrees of freedom in the insulating states 
we derive the effective Hamiltonian of the electronic spins by the perturbative approach. 
We start from the strong coupling limit, $U/t, V/t, U/J, V/J \rightarrow \infty$, where 
the charges localize on every other site as shown in Figs.~\ref{f4}(a) and \ref{f4}(b). 
At the first-order perturbative level of $J$, 
the ${\cal H}_{\rm Ising}$-term in Eq.(\ref{ham}) works as an effective internal magnetic field, $H_{\rm eff}$, 
on these electronic spins. 
Then the system is understood as the non-interacting SU(2) spins under the (internal) magnetic field. 
The spin-exchange interactions within the electronic chain appear 
in the perturbation processes at the fourth order level in terms of $t/U, t/V$. 
Figure~\ref{f4}(c) shows the process which mix the adjacent spins (by two lattice spacing). 
Then, ${\cal H}_{\rm el}$ in Eq.(\ref{elham}) is transformed 
to the effective Hamiltonian given as, 
\begin{eqnarray}
{\cal H}_{\rm eff}\!\!&=&\hspace*{-5mm}
\sum_{j=2l,\;(l={\rm integer})} \hspace*{-5mm}
\Big(\;
J_{\rm eff}^z s_j^z s_{j+2}^z +J_{\rm eff}^\perp (s_j^x s_{j+2}^x + s_j^ys_{j+2}^y) 
\nonumber\\
&& \hspace{2.5cm} -s_j^z H_{\rm eff}(l). 
\;\Big), 
\label{effham}
\end{eqnarray}
Here, $H_{\rm eff}$ is a $l$-dependent internal magnetic field from the localized Ising moments. 
For the ($\up\dwn\up\dwn$)-state we have, 
\begin{eqnarray}
&&J_{\rm eff}^z = \frac{2t^4}{U}\left(\frac{1}{(V+J/2)^2}+\frac{1}{(V-J/2)^2}\right),
\nonumber \\
&& J_{\rm eff}^\perp = \frac{4t^4}{U}\frac{1}{V^2-(J/2)^2}, 
\nonumber \\
&& H_{\rm eff}= \frac{J}{2}
\label{jeff1010}
\end{eqnarray}
where $J_{\rm eff}^z >J_{\rm eff}^\perp$. 
Therefore, the system is {\it an XXZ-spin system with Ising anisotropy in the uniform magnetic field}. 
When finite $J$ is introduced the Ising gap opens, which, however, is suppressed 
immediately by the magnetic field when $J$ further increases. 
We show the evaluated $J_{\rm eff}$'s as a function of $J/t$ in Fig.~\ref{f3}(g) 
from Eq.(\ref{jeff1010}) at $U/t=8,V/t=4$. 
Actually, the increase of $J_{\rm eff}$ is slower than that of $H_{\rm eff}$. 
Therefore, even though the anisotropy of interaction, $J_{\rm eff}^z / J_{\rm eff}^\perp$, 
increases as a function of $J$, the uniform magnetic field $H_{\rm eff}$ has larger magnitude and 
the spin sector is gapless. 
\par
On the other hand, the ($\up\up\dwn\dwn$)-Ising spin configuration yields the 
following effective parameters, 
\begin{eqnarray}
&&J_{\rm eff}^z = t^4\bigg(  \Big(\frac{1}{V^2}+\frac{1}{(V+J/2)^2}\Big)\frac{1}{U+J/2}\bigg)
\nonumber\\
 && \hspace{20mm}+\Big(\frac{1}{V^2}+\frac{1}{(V-J/2)^2}\Big)\frac{1}{U-J/2},
\nonumber \\
&& J_{\rm eff}^\perp = t^4 \bigg(
\frac{1}{V}\Big(\frac{1}{(V+J)(U+J/2)}+\frac{1}{(V-J)(U-J/2)}\Big) 
\nonumber\\
 && \hspace{15mm}
+\frac{1}{(V+J/2)^2(U+J/2)}+\frac{1}{(V-J/2)^2(U-J/2)}\bigg), 
\nonumber \\
&& H_{\rm eff} = (-)^l\frac{1}{2}\bigg( J
 +\Big(\frac{1}{V^2}+\frac{1}{(V+J/2)^2}\Big)\frac{t^4}{U+J/2}
\nonumber\\
 && \hspace{30mm}
  -\Big(\frac{1}{V^2}+\frac{1}{(V-J/2)^2}\Big)\frac{t^4}{U-J/2}
 \bigg)
\label{jeff1100}
\end{eqnarray}
This time we have $J_{\rm eff}^z < J_{\rm eff}^\perp$, 
and the system is interpreted as {\it an XXZ-spin chain with XY-anisotropy (TLL) 
placed under the staggered magnetic field}. 
Figure~\ref{f3}(g) shows the increase of $J_{\rm eff}$ as a function of $J$. 
Again, the internal field overwhelms the effective spin interactions. 
The N\'eel order is stabilized, which has large spin gap compared to the ($\up\dwn\up\dwn$)-state. 
This is simply because the period of correlations of Ising and electronic spins are commensurate, 
and they cooperate by the introduction of $J$. 
The large $\expct{s_z}$ of local moments in the DMRG calculation in Fig.~\ref{f3}(d) 
actually supports this scenario. 
\par
We mention that at $J=0$, the effective Hamiltonians, Eqs.(\ref{jeff1010}) and (\ref{jeff1100}) 
are reduced to the simple Heisenberg spin Hamiltonian without the magnetic field ($H_{\rm eff}=0$), 
which has a SU(2) spin-spin interaction, $J_{\rm eff}^z=J_{\rm eff}^\perp=t^4/(UV^2) \equiv J_{\rm eff}^0$, 
which we mentioned in Sec. II. 
The symmetry of $J_{\rm eff}$ at $J\ne 0$ is thus modified to a Z(2)-one by the localized Ising moments 
of Z(2) symmetry. 
\begin{figure}[t]
\includegraphics[width=8cm]{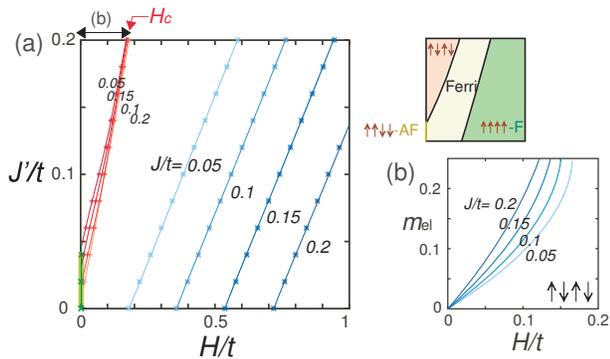}
\caption{DMRG results analyzed by the combination with the external magnetic field ($H$) at $U/t=8$, $V/t=4$ and $N$=64. 
Panel (a) is the phase diagram on the plane of $J'$ and $H$. 
Panel (b) gives the magnetization, $m_{\rm el}$ of the $(\up\dwn\up\dwn)$-phase at the small $H$ 
(the corresponding region is indicated by arrows above the phase diagram). 
} 
\label{f5}
\end{figure} 
\begin{figure}[t]
\includegraphics[width=8cm]{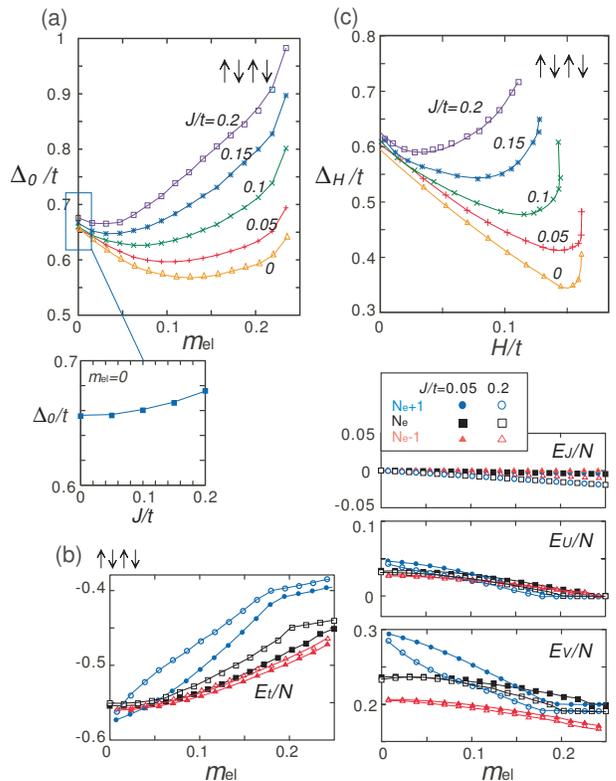}
\caption{
Panel (a) and (c) are "energy gap" at $H=0$ ($\Delta_0$ without the Zeeman term) 
and $H\ne 0$ ($\Delta_H$ including the zeeman term) as a function of $m_{\rm el}$ and $H$, respectively, 
where (c) corresponds to the charge gap under the magnetic field. 
The charge gap ($\Delta_0$ at $m_{\rm el}=0$) at $H=0$ as a function of $J$ is shown together. 
Panel (b) is the energy of the electronic state of $(\up\dwn\up\dwn)$ 
at $N_e(=N/2)$, $N_e+1$, and $N_e-1$ as a function of $m_{\rm el}$. 
$E_U$, $E_V$, $E_J$, and $E_t$ are the energies of $U$, $V$, $J$, and $t$-terms in Eq.(\ref{ham}). 
} 
\label{f6}
\end{figure} 
\section{External magnetic field}
\subsection{Phase diagram under external field}
Finally, we introduce the external magnetic field, $H$, to the present system with rather complicated magnetism. 
Figure~\ref{f5}(a) shows the phase diagram on the plane of $J'/t$ and $H/t$ for several choices of $J$ at $N=64$. 
We consider the configuration of local Ising moments up to 32-site periodicity, 
all possible degrees of electronic polarization, $m_{\rm el}$, and determine the lowest energy state. 
The phase diagram is classified by the configuration of localized moments into four parts; 
at small $J'$ and $H \sim 0$, the ($\up\up\dwn\dwn$)-phase exists, which is immediately 
replaced by the ferrimagnetic phase by the introduction of $H$. 
The stabilized ferrimagnetic states are the periodic ones, 
e.g., those with four fold, ($\up\up\up\dwn$), or eight fold periodicity, ($\up\up\up\up\up\up\up\dwn$), 
and finally the full saturation of magnetic moments on both chains takes place. 
The saturation field (onset of ($\up\up\up\up$)-F state) of localized moments is approximately given as $H_s \sim 4J+2J'$. 
At larger $J'$, the ($\up\dwn\up\dwn$)-phase sustains from $H$=0 to finite magnetic field. 
\par
Here, again the ($\up\up\dwn\dwn$)-and ($\up\dwn\up\dwn$)-states significantly differ regarding the instability against 
the external magnetic field.  
The former state is unstable and disappears at finite $H$. 
This is because it has no energy gain in the Ising part, $E_{\rm Ising}=0$, which allows for 
the flipping of Ising moments into the ferrimagnetic configuration. 
Here, the small spin gap $\sim J/2$ of the electronic spins does not make much advantage. 
In contrast, the ($\up\dwn\up\dwn$)-phase sustains due to energy gain, 
$E_{\rm Ising}=-NJ'/4$, which is the largest among all other configurations. 
In this phase, the Ising moment is stable while the magnetization of electrons, $m_{\rm el}$ gradually changes with $H$. 
One can adopt the magnetization curve in Fig.\ref{f5}(b) to the ($\up\dwn\up\dwn$)-phase in 
Fig.~\ref{f5}(a) along the $H/t$-axis regardless of the value of $J'/t$.
As we saw in Fig.\ref{f3}(a) the $m_{\rm el}$-dependence of the energy of this charge ordered state was relatively small at finite $J$. 
Therefore, the Zeeman term of the electronic spins "absolves" the effect of the external magnetic field 
before the localized spins start to flip at the $J'/t$-dependent $H=H_c$ 
(i.e. the ($\up\dwn\up\dwn$)- and ferrimagnetic phase boundary in Fig.~\ref{f5}(a)). 
%
%
\subsection{Charge gap under the external field}
We discussed in Sec.IV (Fig.~\ref{f3}(c)) that the amplitude of charge ordering in the ($\up\dwn\up\dwn$)-state 
increases by the polarization of electronic spins. 
Therefore, the charge gap is also expected to increase with $m_{\rm el}$. 
We calculate the energy gap $\Delta_o=E(N_e-1)+E(N_e+1)-2E(N_e)$ as a function of $m_{\rm el}$. 
For its evaluation at $(N_e,m_{\rm el}$) with $N_e=N/2$, 
we take either of the magnetic polarization $m_{\rm el}\pm 1$ which gives the lower energy 
in both the electron-doped $(N_e+1)$ and hole-doped $(N_e-1)$ states. 
The result as a function of $m_{\rm el}$ is shown in Fig.~\ref{f6}(a) for several choices of $J$. 
As expected, the fully saturated state has larger $\Delta_0$ than the unpolarized state. 
However, at intermediate $0<m_{\rm el}<0.25$ we find a significant decrease of $\Delta_0$. 
This decrease is several orders of magnitude larger compared to the variation of $\Delta_0$ 
induced by $J$ at $m_{\rm el}$=0 (see the lower panel of Fig.~\ref{f6}(a)), 
and is not attributed to $J$ since the decrease is the largest at $J=0$. 
\par
Let us consider the origin of the particular $m_{\rm el}$-dependence of $\Delta_0$. 
Figure~\ref{f6}(b) shows the interaction and kinetic energies, 
$E_U$, $E_V$, $E_J$, and $E_t$, for ($N_e$)- and ($N_e\pm 1$)-states, 
which all together contribute to $\Delta_0$. 
What we find is the following; focusing on the energy difference 
between different electron numbers, 
one sees that $E_U$ and $E_J$ do not differ much between $(N_e)$- and $(N_e\pm 1)$-states. 
Whilst, in the $(N_e+1)$-state, $E_V$ and $E_t$ have large energy loss at small and large $m_{\rm el}$, respectively, 
compared to $(N_e)$ and $(N_e-1)$-states. 
\par
Here, $E_t$ is an increasing function of $m_{\rm el}$, namely the paramagnetic state has larger kinetic gain compared to the ferromagnetic one. 
This originates from the Pauli's principle as in the usual band picture. 
At large $U$ and $V$, the doping of electron makes the system less itinerant in overall 
since the electrons try to avoid each other at most within the limited spacing. 
Such tendency is more significant for larger $m_{\rm el}$ where there is less space for electrons to share due to the Pauli's principle. 
At the same time, in the partially polarized case, 
the electrons locally follow the configuration realized at $m_{\rm el}=0$ to gain maximally the kinetic energy. 
Thus $E_t$ at $(N_e+1)$ turns out to be a convex downward function. 
There is an energy loss of $E_V$ whenever there is a gain in $E_t$, so that $E_V$ in ($N_e+1$)-state behaves 
contrary to that of $E_t$, namely as a convex upward function. 
Then, these convex functions together suppresses the charge gap at $0<m_{\rm el}<0.25$. 
\par
The convex downward functional form of $\Delta_0$ is the most distinct at $J=0$, 
while its both edges have the comparable values, $\Delta_0(m_{\rm el}=0)\sim \Delta_0(m_{\rm el}=0.25)$. 
When $J$ becomes finite, the kinetic energy gain is significantly suppressed at $m_{\rm el}>0$, 
since the polarized electrons are pinned on every other site by the localized moments. 
Thus, $E_t$ of the ($N_e+1$)-state at $m_{\rm el}>0$ increases with $J$, particularly at 
smaller $m_{\rm el}$ which was originally more itinerant. 
Then, the convex functional form dissolves and $\Delta_0$ is gradually transformed towards a monotonically 
increasing function of $m_{\rm el}$ with increasing $J$. 
\par
Next, we include the Zeeman term and calculate the charge gap against the external magnetic field, 
which is denoted as $\Delta_H$. 
Reflecting the functional form of $\Delta_0$, 
it appears as a convex-downward function as shown in Fig.~\ref{f6}(c). 
Here, for the doped cases we again choose either of the $m_{\rm el}\pm 1$ which gives the lower energy including the Zeeman terms. 
Notice that we neglect the orbital effect under the external field, which we consider to be small in one-dimension. 
Particularly at small $J/t$ the gap continues to decrease significantly towards $H/t \sim 0.1-0,2$. 
Since the flipping of the Ising moments to the ferrimagnetic phase takes place at 
$H_c\sim J'$, the upturn of the gap does not appear in overall in the $(\up\dwn\up\dwn)$-phase in Fig.~\ref{f5}(a). 
\par
In this way, the decrease of $\Delta_0$ by $m_{\rm el}$ characteristic of the charge ordered state is combined with 
the stability of the ($\up\dwn\up\dwn$)-Ising state, 
and together allow for the suppression of charge gap $\Delta_H$ by the magnetic field. 
%
\section{Summary and Discussions}
In the present paper, we disclosed the intriguing interplay of 
magnetic and electric properties of the quarter-filled strongly correlated 
electronic chains coupled to the Ising moments. 
Almost regardless of the details of the electronic state, 
the ground state is in overall classified by the two different configuration 
of Ising moments, ($\up\dwn\up\dwn$) and ($\up\up\dwn\dwn$). 
The latter is an analogue of the paramagnetic state of the Kondo lattice model 
which is stabilized by the RKKY-interaction via Kondo coupling, $J$. 
When the direct interaction between Ising moments, $J'$, is introduced, 
this ($\up\up\dwn\dwn$)-state is replaced by the ($\up\dwn\up\dwn$)-state. 
\par 
The main focus is the interplay of spin and charge degrees of freedom 
in the insulating state with large $U$ and $V$ coupled to the Ising moments. 
The electrons are localized on every other site, which are interacting 
antiferromagnetically via $J_{\rm eff}$ within the chain. 
The Ising moments work as effective internal field $|H_{\rm eff}|=J/2$ 
to these electronic spins. 
At the same time the Ising moments modify the symmetry of the interaction, $J_{\rm eff}$, 
from the SU(2) at $J$=0 to the Z(2) ones. 
Thus, the electronic spins coupled to the ($\up\dwn\up\dwn$)-Ising moments 
behave as an Ising XXZ-spin system under uniform magnetic field, 
and the one coupled to the ($\up\up\dwn\dwn$)-Ising moments 
is regarded as a XY-spin system under staggerd magnetic field. 
Since $H_{\rm eff}$ is higher in order than the Z(2)-modification of interactions, 
the electronic spins are gapless($\up\dwn\up\dwn$) and gapped($\up\up\dwn\dwn$), respectively. 
\par
The more simple description is given in the following; 
when the Ising and electronic chains are decoupled, 
the period of magnetic correlation of the electronic chain 
differs by twice from ($\up\dwn\up\dwn$) while the same as ($\up\up\dwn\dwn$). 
In the former case, the magnetic incommensurability of the two chains 
leads to the suppression of the antiferromagnetic correlation of electronic spins. 
In such case, the $m_{\rm el}$-dependence of energy becomes small, so that 
the external field successively flips the electronic spins by the Zeeman effect, 
while the ($\up\dwn\up\dwn$)-Ising order sustains due to the large energy gain by $J'$. 
Along with the polarization of the electronic spins, the charge gap decreases significantly by $H$. 
Contrastingly, the ($\up\up\dwn\dwn$)-configuration is stabilized solely 
by the magnetic commensurability of two chains (and not by $J'$), 
thus is easily destroyed by the external magnetic field. 
We conclude that the ($\up\dwn\up\dwn$)-order of Ising spins 
appear as a consequence of the direct interaction between Ising moments $J'$, 
where a frustration of competing magnetic orders is embedded. 
Such frustration yields a strong-correlation-driven interplay of spin and charge degrees of freedom. 
\par
The similar picture shall also be found in the EKLM. 
The paramagnetic state of EKLM corresponds to ($\up\up\dwn\dwn$) in the present study. 
The ($\up\up\dwn\dwn$)-order is weakened in the EKLM 
by the quantum fluctuation and transformed into the antiferromagnetic 
correlation which cooperates with $J_{\rm eff}$ via RKKY interaction\cite{tsunetsugu93}. 
When $J$ becomes large, the SU(2) spins form singlets with the electrons, 
which propagate and stabilize the ferromagnetism\cite{tsunetsugu93}. 
Therefore, if one includes the direct interaction, $J'$, between SU(2) spins, 
it favors antiferromagnetism and competes with the $J$-induced ferromagnetism. 
Thus, a similar magnetic frustration may appear. 
However, such physical picture shall be rather blurred by the quantum fluctuation(SU(2)). 
The present Ising spin system has a more serious frustration effect, 
which may lead the electronic system to a sensitive response against the magnetic field. 
\par
Finally, let us examine the relevance of these results with the experiments on 
the TPP[$M$(Pc)(CN)$_2$]$_2$, $M$=Fe,Co\cite{tajima_chem}. 
The Co-salt is a pure electronic chain ($J=0$ in the present model), which is a good 
reference system to analyze the effect of localized moments. 
Both salts have semi-conducting temperature dependence of resistivity. 
The Fe-salt shows a large negative magnetoresistance (MR)
which amounts to $\rho(H)/\rho(0)\sim 10^{-2}$, where $\rho(H)$ is the resistivity under the 
magnetic field $H$. 
The experimental findings are summarized as follows; 
(1) the activation energy derived from the resistivity data are $\Delta_a\!\sim\!10^{-3}$eV 
and $\!\sim\!10^{-2}$eV for Co-and Fe-salts, respectively\cite{hanasaki00}. 
(2) The localized moment has anisotropic g-values, 
$g_\perp\sim 3.6$ and $g_{\parallel}\sim 0.5-1$ which are 
roughly perpendicular and parallel to the molecular-axis, respectively. 
This anisotropy is explained in terms of the spin-orbit coupling and the resultant 
magnetic moment is $S=1/2$ \cite{hanasaki03}. 
(3) The magnetic susceptibility shows a large anisotropy, $\chi_{\perp}/\chi_{\parallel} > 5$, 
and $\chi_{\parallel}$ gives similar values with the Co-one\cite{hanasaki00}. 
(4) Residual magnetization is observed at $T<12$K, which is attributed to the ferrimagnetism of $\pi$-electrons 
by the torque experiment\cite{tajima08}. 
(5) The MR is not scaled by the magnetization and 
shows large $T$-dependence\cite{hanasaki06}. 
(6) Phase transition is absent (in contrast to DEX), 
namely the scenario of the competition of two different orders are not applicable. 
(7) The MR is relevant when Fe-ion is partially replaced by Co-ion, e.g. 
even when the degree of replacement is as large as Fe$_{0.07}$Co$_{0.93}$, the MR amounts to $\rho(H)/\rho(0)\sim 0.5$\cite{hanasaki06-2}, 
(8) the magnetization of electrons gradually increases with $H$, and that of the Fe-moments starts at $H_c\sim$ 15T
\cite{hanasaki_pr}. 
(9) the ground state of the electronic chain is weakly charge ordered which is observed 
by the NQR study\cite{hanasaki06}. 
\par
The factor (2) and (9) are taken into account in the present model. 
The parameter values estimated from the extended H\"uckel calculation, ab-initio calculation and 
by the reflectance spectra give $t\sim$0.1eV, $J\sim t/10$, and $J'\sim J/3$, 
where $J$ and $J'$ are Hund and exchange coupling constants, respectively. 
If we put these parameters onto the phase diagram of Fig.~\ref{f2}(b), 
it is located within the $(\up\dwn\up\dwn)$-phase. 
The magnetic properties of electrons behave quite sensitive to $U$ and $V$; 
when $U=8$ and $V=4$, we see that the system is in the vicinity of the phase boundary between the AF- and F-electrons. 
As we saw in Fig.~\ref{f3}(a) the $m_{\rm el}$-dependence of energy is small. 
Then, the ferrimagnetism of $\pi$-electrons is possible, which is consistent with (4). 
The external field gradually magnetizes the $\pi$-electrons, 
and the onset of flipping of localized moments, $H_c\sim$ 0.01$t$ in (8) is consistent with the phase digram in Fig.~\ref{f5}(a) (at $J'/t \!\sim \!0.1/3$). 
In this way, we reach the picture that electrons show fragile magnetic properties 
under the stable antiferromagnetism of localized Ising moment supported by $J'$. 
\par
However, there still remains some issues to be clarified; 
in our model, the charge gap decreases with $H$, while it does decrease even at $J=0$. 
Therefore, $J$ does not seem to favor the suppression of charge order by the magnetic field. 
While, the Co-salt shows a small but positive MR\cite{hanasaki06}, which is incompatible with our results. 
Also the present model which focuses on the particular density of d-spins 
cannot cope with the issue (7). 
Therefore, in order to clarify fully the origin of magnetoresistance, a more systematic experimental data 
(regarding the Fe-Co-ratio or the effect of the dimensionality of the system), as well as 
the theoretical calculations directly on the transport properties shall be required. 
\par
The author acknowledge N. Hanasaki and M. Takigawa for helpful discussions and comments. 
This work is supported by Grant-in-Aid for Scientific 
Research (No. 21110522, 19740218) from the Ministry of Education, Science, Sports and Culture of Japan. 

\end{document}